\documentclass[conference]{IEEEtran}
\IEEEoverridecommandlockouts
\usepackage{cite}
\usepackage{amsmath,amssymb,amsfonts}
\usepackage{algorithmic}
\usepackage{graphicx}
\usepackage{textcomp}
\usepackage{xcolor}
\usepackage{hyperref}
\usepackage{placeins}
\def\BibTeX{{\rm B\kern-.05em{\sc i\kern-.025em b}\kern-.08em
    T\kern-.1667em\lower.7ex\hbox{E}\kern-.125emX}}

\usepackage{siunitx}
\begin{document}

\title{ShuffleUNet: Super resolution of diffusion-weighted MRIs using deep learning}

\author{\IEEEauthorblockN{Soumick~Chatterjee\textsuperscript{\textsection}\IEEEauthorrefmark{1}\IEEEauthorrefmark{2}\IEEEauthorrefmark{3}, 
                         Alessandro~Sciarra\textsuperscript{\textsection}\IEEEauthorrefmark{3}\IEEEauthorrefmark{4}, 
                         Max~D{\"u}nnwald\IEEEauthorrefmark{2}\IEEEauthorrefmark{4},\\
                         Raghava~Vinaykanth~Mushunuri\IEEEauthorrefmark{2},
                         Ranadheer~Podishetti\IEEEauthorrefmark{2},
                         Rajatha~Nagaraja~Rao\IEEEauthorrefmark{2},
                         Geetha~Doddapaneni~Gopinath\IEEEauthorrefmark{2},\\
                         Steffen~Oeltze-Jafra\IEEEauthorrefmark{4}\IEEEauthorrefmark{5}\IEEEauthorrefmark{6},
                         Oliver~Speck\IEEEauthorrefmark{3}\IEEEauthorrefmark{5}\IEEEauthorrefmark{6}\IEEEauthorrefmark{7} and 
                         Andreas~N{\"u}rnberger\IEEEauthorrefmark{1}\IEEEauthorrefmark{2}\IEEEauthorrefmark{6}}

\IEEEauthorblockA{\IEEEauthorrefmark{1}Data and Knowledge Engineering Group, Otto von Guericke University Magdeburg, Germany}
\IEEEauthorblockA{\IEEEauthorrefmark{2}Faculty of Computer Science, Otto von Guericke University Magdeburg, Germany}
\IEEEauthorblockA{\IEEEauthorrefmark{3}Department of Biomedical Magnetic Resonance, Otto von Guericke University Magdeburg, Germany}
\IEEEauthorblockA{\IEEEauthorrefmark{4}MedDigit, Department of Neurology, Medical Faculty, University Hospital Magdeburg, Germany}
\IEEEauthorblockA{\IEEEauthorrefmark{5}German Centre for Neurodegenerative Diseases, Magdeburg, Germany}
\IEEEauthorblockA{\IEEEauthorrefmark{6}Center for Behavioral Brain Sciences, Magdeburg, Germany}
\IEEEauthorblockA{\IEEEauthorrefmark{7}Leibniz Institute for Neurobiology, Magdeburg, Germany}

}

\maketitle

\begingroup\renewcommand\thefootnote{\textsection}
\footnotetext{S. Chatterjee and A. Sciarra contributed equally}
\endgroup

\begin{abstract}
Diffusion-weighted magnetic resonance imaging (DW-MRI) can be used to characterise the microstructure of the nervous tissue, e.g. to delineate brain white matter connections in a non-invasive manner via fibre tracking. Magnetic Resonance Imaging (MRI) in high spatial resolution would play an important role in visualising such fibre tracts in a superior manner. However, obtaining an image of such resolution comes at the expense of longer scan time. Longer scan time can be associated with the increase of motion artefacts, due to the patient's psychological and physical conditions. Single Image Super-Resolution (SISR), a technique aimed to obtain high-resolution (HR) details from one single low-resolution (LR) input image, achieved with Deep Learning, is the focus of this study. Compared to interpolation techniques or sparse-coding algorithms, deep learning extracts prior knowledge from big datasets and produces superior MRI images from the low-resolution counterparts. In this research, a deep learning based super-resolution technique is proposed and has been applied for DW-MRI. Images from the IXI dataset have been used as the ground-truth and were artificially downsampled to simulate the low-resolution images. The proposed method has shown statistically significant improvement over the baselines and achieved an SSIM of $0.913\pm0.045$.  
\end{abstract}


\section{Introduction}
Non-invasive brain imaging techniques have been advancing for the past few decades and are used for detecting various diseases, to study brain anatomy and its functions~\cite{despotovic2015mri}. Diffusion tensor imaging (DTI) or diffusion-weighted imaging (DWI) is one of the MR techniques, which is employed in diagnosing white matter diseases, cancer or stroke. The proportion of water in the human body is approximately $60-70\%$ and it is distributed over intra- and extracellular compartments. Different tissues in the human body exhibit varying diffusion characteristics. DWI is a technique, which generates signals based on changes in Brownian motion and it provides information regarding the diffusion properties. Axon membranes present in brain white matter limit the molecular movement perpendicular to the fibre thus resulting in anisotropic diffusion in white matter. DWI uses this property to provide information on white matter integrity and structural details of white matter tracts~\cite{le2014diffusion,baliyan2016diffusion}. Though DWI is advancing rapidly and used widely in medical diagnosis, obtaining high-resolution scans is rather difficult since it requires longer acquisition times, which may lead to motion artefacts. Single Image Super-Resolution (SISR), a technique aimed to obtain high-resolution (HR) details from one single low-resolution (LR) input image, is the focus of this study. 

\subsection{Related Work}
Deep learning has emerged in recent times as one of the most important tools for various applications, and SISR is one of the hot topics in the world of deep learning. Various deep learning based SISR techniques have been developed over the years~\cite{zeng2018simultaneous,he2020super}. UNet~\cite{ronneberger2015u,cciccek20163d}, originally proposed for image segmentation, has gained popularity in a wide array of applications since its inception, including as a solution of inverse problems such as SISR of MRI reconstruction~\cite{iqbal2019super,sarasaen2021fine}. Han et al.~\cite{han2018framing} proposed a tight-frame UNet architecture exploiting wavelet decomposition, which improves the performance of UNet for inverse problems (shown for 2D sparse-view CT). Even though UNet (and its variants) is one of the most common architectures for the task of SISR and other inverse problems, it has its limitations, such as smoothed output and checker-board artefacts. Aitken et. al~\cite{aitken2017checkerboard} attributed such problems to the following components of UNet - deconvolution layer for upsampling, strided convolutions or max-pooling for downsampling and random weight initialisation. It has been observed that upsampling from low-resolution feature maps to higher resolution, commonly referred to as deconvolution step, can produce checker-board patterns. Sub-pixel convolution~\cite{Shi_2016_CVPR} is observed to have an upper hand over the other methods in avoiding such artefacts~\cite{aitken2017checkerboard}. Max-pooling is conventionally used for downsampling operations in UNet-based models, which chooses the maximum value in the given kernel and there is a loss of pixel information which is not desirable as there can be a high possibility for loss of sensitive information. Toutounchi et al.~\cite{toutounchi2019advanced} proposed a lossless pooling layer to deal with the blurring problem (loss of information) of the max-pool operation. Furthermore, it has been shown that batch normalisation layers (part of the UNet architecture) can negatively impact the result as they restrict the range flexibility of the networks by normalising the features and also increase the GPU memory requirements by around 40\%~\cite{nah2017deep,lim2017enhanced}. One final point to be noted is that typically deep learning based techniques require large training datasets. Employing patch-based techniques (such as patch-based super-resolution: PBSR) might be able to deal with the problem of limited training data~\cite{jain2017patchSR}, while also reducing the GPU memory requirements. 

\subsection{Contribution}
This research work proposes ShuffleUNet architecture for 3D volumetric images, inspired by tight frame UNet, which addresses the aforementioned problems of blurred or smoothed output and checker-board artefacts by replacing the strided convolutions with lossless pooling layers (pixel unshuffle), by replacing deconvolution (transposed convolution or interpolation) with a sub-pixel convolution operation (pixel shuffle) and by removing the batch normalisation layers. This paper validates the proposed model by performing patch-based single-image super-resolution of diffusion-weighted images and shows the advantage of such an approach in the visualisation of fibre tracts and other derived data.

\section{Methodology}

\subsection{Network Architecture}
\begin{figure*}[htbp]
    \centering
    \includegraphics[width=\textwidth]{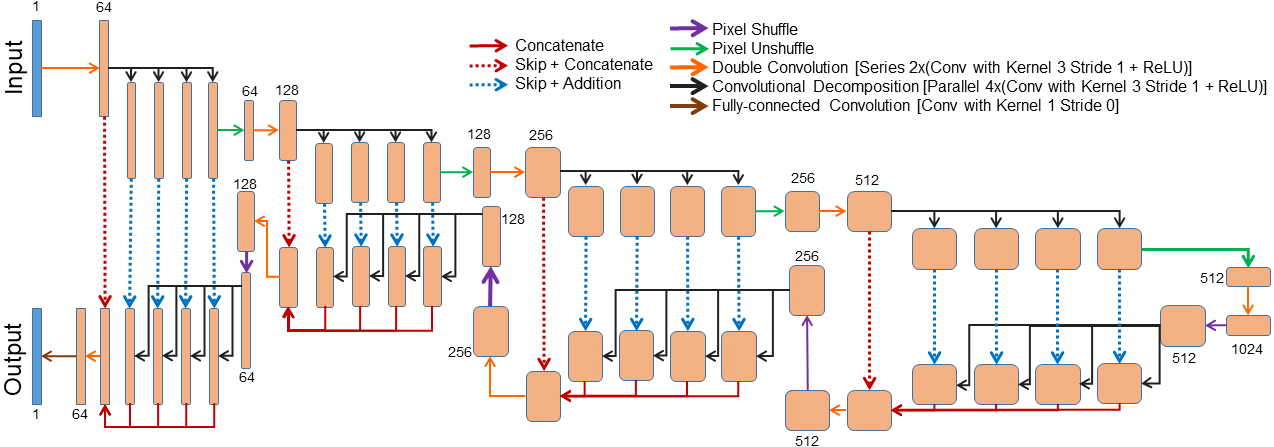}
    \caption{Proposed network architecture: ShuffleUNet}
    \label{fig:net}
\end{figure*}

The proposed ShuffleUNet consists of four blocks in the contraction path, each of them downsamples the input by half in all dimensions. Each block consists of three sub-blocks: double convolution, convolutional decomposition, and pixel unshuffle. The input goes to double convolution and its output serves as the input of the convolutional decomposition sub-block. The input of this sub-block is provided to each convolution of this block and four different outputs are obtained - these outputs are referred to as convolutional decomposition of the input of this sub-block. Pseudo-loss-less downsampling operation, pixel unshuffle (see Eq.\ref{eq:PUS}), is applied on the fourth output, which down-samples the input by a factor of two in all dimensions, and the rest of the outputs are directly forwarded as skip-connection to the expansion path.
\begin{equation}
\label{eq:PUS}
I ^D = f^l(I ^L) =  \textbf{\emph{PU}}(W ^l \times f ^{(l-1)}(I ^L) + b ^l)
\end{equation}
Where \textbf{\emph{PU}} denotes the pixel-unshuffling operation, which rearranges all the elements of a tensor of dimension (n, c, rH, rW, rD) tensor to a tensor of dimension (n, \(r^3\) x c, H , W , D), r is the scaling factor. After the contraction path, one double convolution sub-block is applied to the output of the final pixel unshuffle as the latent convolution. The output of this latent convolution is passed to the expansion path. 

Similar to the contraction path, the expansion path also contains four blocks - each of them upsamples the input by a factor of two in all dimensions. Each of these blocks contains three sub-blocks: pixel shuffle, convolutional decomposition, and double convolution. Pixel shuffle sub-blocks upscale the given input by a factor of k (here k=2) in all dimensions by reducing the number of filters by a factor of 1/\(k^3\). A tensor of dimensions (n, \(r^3\) x c, H, W, D) is upsampled with periodic shuffling operation in feature space as explained by the following equation:
\begin{equation}
\label{eq:PS}
I ^S = f^l(I ^L) =  \textbf{\emph{PS}}(W ^l \times f ^{(l-1)}(I ^L) + b ^l)
\end{equation}
where \textbf{\emph{PS}} denotes pixel-shuffling operation which shuffles and arranges all the pixel elements of a tensor with dimensions (n, \(k^3\) x c, H, W, D) to a tensor with dimensions (n, c, kH × kW x kD), k is the scaling factor. \(W ^l\), \(b^l\) denotes the weights and bias matrices and \(f ^{(l-1)}\) represents feature maps from low-resolution feature space from the previous layer and \(f^l\) represents upsampled feature space after pixel shuffling. The output of the pixel shuffle sub-block is forwarded to the convolutional decomposition sub-block to obtain four different outputs which are then added with the incoming skip-connections from the same level of the contraction path, and finally, these four results are concatenated together. Then, this is further concatenated together with the output of the skip-connection coming from the pixel unshuffle operations of the contraction path, which is then forwarded to the double convolution sub-block. The final output of the model is obtained after the fully-connected convolution layer. 

The initial number of filters of the network (first convolution layer) is $64$ and the number of features is doubled by each of the contraction path blocks after every down-sampling step and is reduced by half in every up-sampling process in the expansion path. Also, it is noteworthy that here both pixel shuffle and pixel unshuffle operators have learnable parameters. Furthermore, the weight initialisation (convolution, pixel shuffle and pixel unshuffle layers) was performed using a normal distribution (also known as Kaiming Normal)~\cite{he2015delving}, rather than the typical choice of uniform distribution (Kaiming Uniform), to help the convergence and to avoid degrading effects arising from random weight initialisation~\cite{aitken2017checkerboard}. 

\subsection{Implementation}
The implementation was done using PyTorch and was trained using an Nvidia V100 GPU. The volumes were divided into patches with a patch size of $96\times96\times48$ with the help of TorchIO~\cite{perez2020torchio}. The network was trained for 80 epochs (when it converged) with a batch size of four. The loss was calculated using L1 loss (mean absolute error) and was optimised using the Adam optimiser with a learning rate of \num{1e-4}.

\subsection{Evaluation}
The results of the proposed model and the baselines (Sinc interpolation, trilinear interpolation and UNet model) were compared to the ground truth data using three metrics: structural similarity index (SSIM)~\cite{wang2004imageSSIM}, root-mean-square error (RMSE), and universal quality index (UQI)~\cite{wang2002universal}. Furthermore, derived voxel-wise data, such as axial diffusivity (AD), fractional anisotropy (FA), mean diffusivity (MD), and diffusion tensors (E)~\cite{solowij2017chronic}, were obtained using DIPY~\cite{garyfallidis2014dipy}. These derived data can be utilised to generate the fibre tracts and assess the integrity of white-matter microstructure~\cite{solowij2017chronic}. The derived data from the ShuffleUNet results and baseline results were compared against the derived data obtained from the ground-truth volumes quantitatively using RMSE and UQI.

\subsection{Dataset}
For this research, DWIs from the IXI dataset \footnote{IXI Dataset: \url{https://brain-development.org/ixi-dataset/}} were used, which were acquired with 1.5T and 3T MRI scanners with $16$ diffusion directions for $399$ healthy subjects. The volumes in the dataset are originally of anisotropic spatial resolutions $1.75\times1.75\times2.35$ $mm^3$ and $1.75\times1.75\times2$ $mm^3$. The 3D volumes were under-sampled to simulate the low-resolution dataset, by a factor of two in all dimensions using FSL~\cite{jenkinson2012fsl} to a spatial resolution of $3.5\times3.5\times4.7$ $mm^3$, $3.5\times3.5\times4.0$ $mm^3$ respectively, which is $12.5\%$ of the original data. UNet based models require the same size for input and output, hence the low-resolution dataset was then interpolated using sinc interpolation to achieve the same pixel dimensions as the high-resolution volumes before supplying as input. The dataset was split into three subsets: training, test and validation with $300$, $50$ and $49$ subjects respectively. 

\section{Results}
The results have been quantitatively and qualitatively compared initially using the actual structural output of the models and then by comparing the derived data obtained. 

\subsection{Output Evaluation}
\begin{figure}[htbp]
    \centering
    \includegraphics[width=0.48\textwidth]{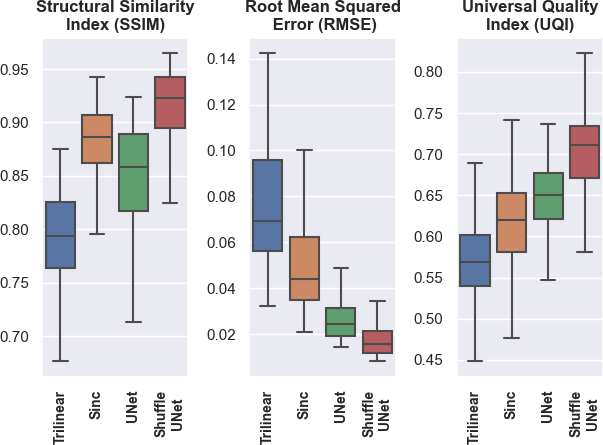}
    \caption{Metrics comparison of the baseline methods and the proposed ShuffleUNet}
    \label{fig:metrics_struct}
\end{figure}

Fig.~\ref{fig:metrics_struct} portrays the resultant SSIM, RMSE and UQI values for the three baselines (trilinear, sinc and UNet) and for ShuffleUNet, while comparing the structural results against the high-resolution ground-truth volumes. It can be observed that trilinear resulted in the lowest values for all three metrics ($0.788\pm0.052$, $0.076\pm0.024$, $0.568\pm0.049$), while ShuffleUNet outperformed all the baselines ($0.913\pm0.045$, $0.017\pm0.007$, $0.702\pm0.060$). It is to be noted that for two of the metrics (RMSE and UQI), UNet achieved the second position after ShuffleUNet, but for SSIM, UNet achieved the third position behind sinc interpolation. 

\subsection{Evaluation of the Derived Data}

\begin{figure*}[htbp]
    \centering
    \includegraphics[width=\textwidth]{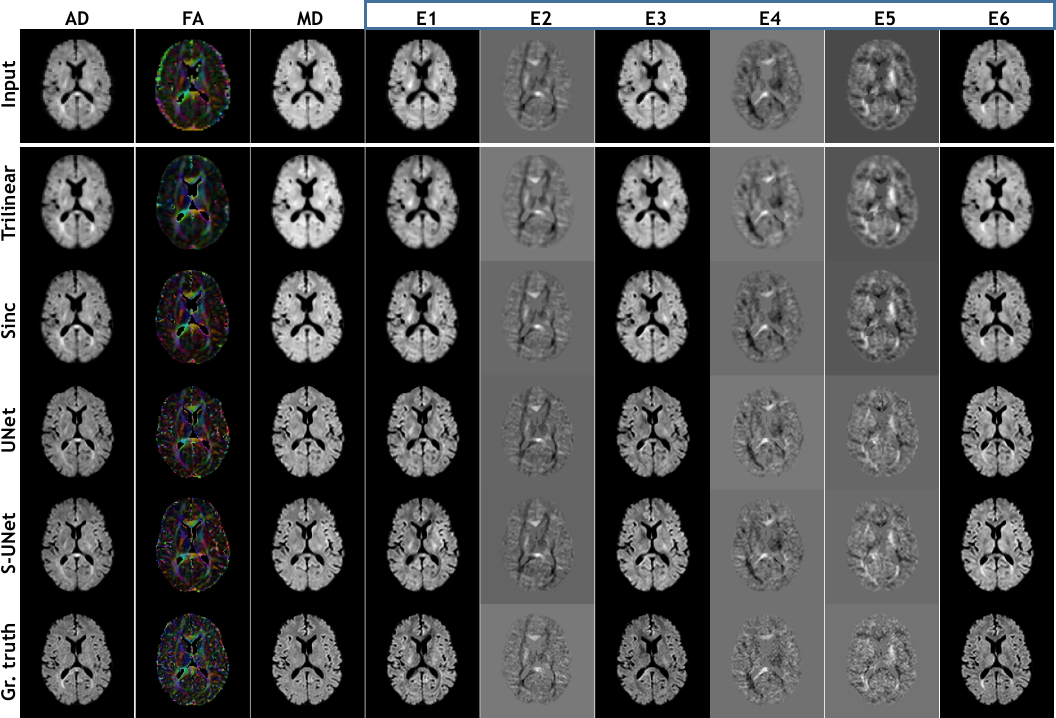}
    \caption{Comparison of derived data (ShuffleUNet, baseline methods and ground-truth). AD: Axial Diffusivity, FA: Fractional Anisotropy, MD: Mean Diffusivity, and E1 to E6: Diffusion Tensors}
    \label{fig:img}
\end{figure*}

\begin{figure}[htbp]
    \centering
    \includegraphics[width=0.48\textwidth]{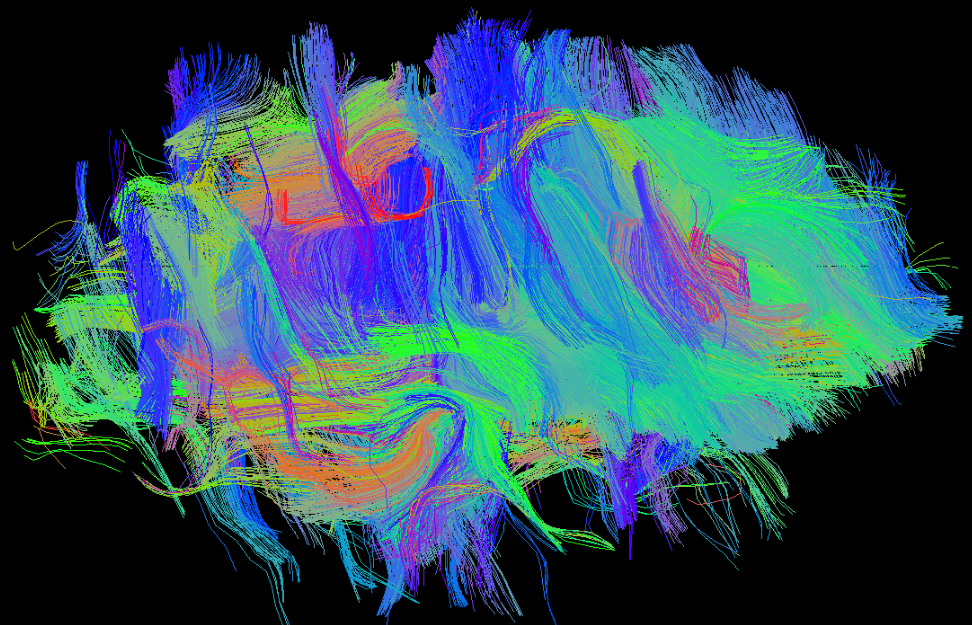}
    \caption{Three dimensional visualisation of the fibre tracts, generated from the derived data obtained from a ShuffleUNet result}
    \label{fig:fibre}
\end{figure}


\begin{figure}[htbp]
    \centering
    \includegraphics[width=0.49\textwidth]{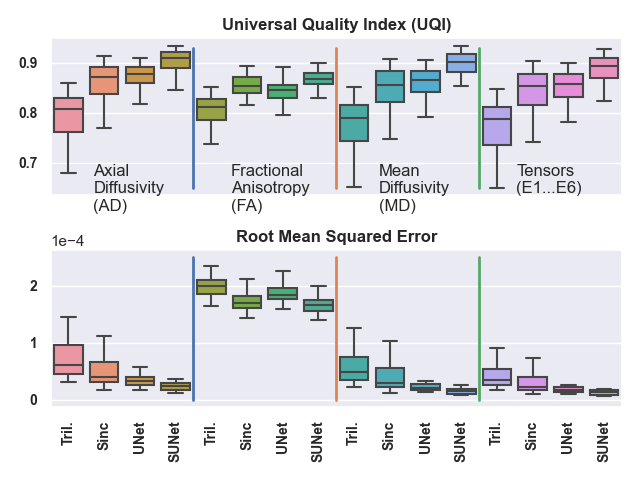}
    \caption{Quantitative evaluation of the derived data obtained from the ShuffleUNet and the baseline methods}
    \label{fig:metrics_derived}
\end{figure}

Fig.~\ref{fig:img} shows the results of the derived data for qualitative analysis and Fig.~\ref{fig:metrics_derived} shows the resultant UQI and RMSE while comparing the derived data obtained from the baseline and ShuffleUNet results against the ground-truth. It can be observed from both metrics that the proposed ShuffleUNet outperformed all the baselines for all four types of derived data. Moreover, Fig.~\ref{fig:fibre} shows an example of obtained fibre tracks from a ShuffleUNet result. 

\subsection{Statistical Hypothesis Testing}
The significance of the improvements observed by the ShuffleUNet were analysed using the independent two-sample t-test, on the basis of the resultant UQI values. Initially, the structural output of the methods was compared and then, the derived data. The resultant p-values obtained while comparing the UQI values of ShuffleUNet against the values of the three baseline methods were always $< 0.01$. Hence, it can be said that ShuffleUNet achieved significant (p-values $<0.05$) improvements over all the baseline methods explored here - for the structural output and also the derived data.

\section{Conclusion}
This paper presents a modified tight-frame UNet architecture, ShuffleUNet, by incorporating pixel unshuffle and pixel shuffle operations for improved down- and upsampling capabilities of the network, thereby reducing artefacts introduced by UNet-based processing pipelines. The method has been evaluated for the task of patch-based single-image super-resolution of diffusion-weighted MRIs. Images from the IXI dataset were downsampled by a factor of two in all dimensions, resulting in a theoretical MRI acceleration factor of eight. The proposed model successfully super-resolves such low-resolution images and achieved statistically significant improvements over the baselines. The evaluation metrics obtained on the derived data indicate that after super-resolving the artificially downsampled images, similar fibre tracts can be derived as the ground-truth images. Hence, images can be acquired with those downsampled resolutions, which will reduce the scan time.  

\section*{Acknowledgement}

This work was partially conducted within the context of the International Graduate School MEMoRIAL at Otto von Guericke University (OVGU) Magdeburg, Germany, kindly supported by the European Structural and Investment Funds (ESF) under the programme "Sachsen-Anhalt WISSENSCHAFT Internationalisierung" (project no. ZS/2016/08/80646). This work was also partially conducted within the context of the Initial Training Network program, HiMR, funded by the FP7 Marie Curie Actions of the European Commission, grant number FP7-PEOPLE-2012-ITN-316716, and supported by the NIH grant number 1R01-DA021146, and by the State of Saxony-Anhalt under grant number 'I 88'.

\bibliographystyle{IEEEtran}
\bibliography{refs} 

\begin{thebibliography}{10}
\providecommand{\url}[1]{#1}
\csname url@samestyle\endcsname
\providecommand{\newblock}{\relax}
\providecommand{\bibinfo}[2]{#2}
\providecommand{\BIBentrySTDinterwordspacing}{\spaceskip=0pt\relax}
\providecommand{\BIBentryALTinterwordstretchfactor}{4}
\providecommand{\BIBentryALTinterwordspacing}{\spaceskip=\fontdimen2\font plus
\BIBentryALTinterwordstretchfactor\fontdimen3\font minus
  \fontdimen4\font\relax}
\providecommand{\BIBforeignlanguage}[2]{{%
\expandafter\ifx\csname l@#1\endcsname\relax
\typeout{** WARNING: IEEEtran.bst: No hyphenation pattern has been}%
\typeout{** loaded for the language `#1'. Using the pattern for}%
\typeout{** the default language instead.}%
\else
\language=\csname l@#1\endcsname
\fi
#2}}
\providecommand{\BIBdecl}{\relax}
\BIBdecl

\bibitem{despotovic2015mri}
I.~Despotovi{\'c}, B.~Goossens, and W.~Philips, ``Mri segmentation of the human
  brain: challenges, methods, and applications,'' \emph{Computational and
  mathematical methods in medicine}, vol. 2015, 2015.

\bibitem{le2014diffusion}
D.~Le~Bihan, ``Diffusion mri: what water tells us about the brain,'' \emph{EMBO
  molecular medicine}, vol.~6, no.~5, pp. 569--573, 2014.

\bibitem{baliyan2016diffusion}
V.~Baliyan, C.~J. Das, R.~Sharma, and A.~K. Gupta, ``Diffusion weighted
  imaging: technique and applications,'' \emph{World journal of radiology},
  vol.~8, no.~9, p. 785, 2016.

\bibitem{zeng2018simultaneous}
K.~Zeng, H.~Zheng, C.~Cai, Y.~Yang, K.~Zhang, and Z.~Chen, ``Simultaneous
  single-and multi-contrast super-resolution for brain mri images based on a
  convolutional neural network,'' \emph{Computers in biology and medicine},
  vol.~99, pp. 133--141, 2018.

\bibitem{he2020super}
X.~He, Y.~Lei, Y.~Fu, H.~Mao, W.~J. Curran, T.~Liu, and X.~Yang,
  ``Super-resolution magnetic resonance imaging reconstruction using deep
  attention networks,'' in \emph{Medical Imaging 2020: Image Processing}, vol.
  11313.\hskip 1em plus 0.5em minus 0.4em\relax International Society for
  Optics and Photonics, 2020, p. 113132J.

\bibitem{ronneberger2015u}
O.~Ronneberger, P.~Fischer, and T.~Brox, ``U-net: Convolutional networks for
  biomedical image segmentation,'' in \emph{International Conference on Medical
  image computing and computer-assisted intervention}.\hskip 1em plus 0.5em
  minus 0.4em\relax Springer, 2015, pp. 234--241.

\bibitem{cciccek20163d}
{\"O}.~{\c{C}}i{\c{c}}ek, A.~Abdulkadir, S.~S. Lienkamp, T.~Brox, and
  O.~Ronneberger, ``3d u-net: learning dense volumetric segmentation from
  sparse annotation,'' in \emph{International conference on medical image
  computing and computer-assisted intervention}.\hskip 1em plus 0.5em minus
  0.4em\relax Springer, 2016, pp. 424--432.

\bibitem{iqbal2019super}
Z.~Iqbal, D.~Nguyen, G.~Hangel, S.~Motyka, W.~Bogner, and S.~Jiang,
  ``Super-resolution 1h magnetic resonance spectroscopic imaging utilizing deep
  learning,'' \emph{Frontiers in oncology}, vol.~9, 2019.

\bibitem{sarasaen2021fine}
C.~Sarasaen, S.~Chatterjee, M.~Breitkopf, G.~Rose, A.~N{\"u}rnberger, and
  O.~Speck, ``Fine-tuning deep learning model parameters for improved
  super-resolution of dynamic mri with prior-knowledge,'' \emph{arXiv preprint
  arXiv:2102.02711}, 2021.

\bibitem{han2018framing}
Y.~Han and J.~C. Ye, ``Framing u-net via deep convolutional framelets:
  Application to sparse-view ct,'' \emph{IEEE transactions on medical imaging},
  vol.~37, no.~6, pp. 1418--1429, 2018.

\bibitem{aitken2017checkerboard}
A.~Aitken, C.~Ledig, L.~Theis, J.~Caballero, Z.~Wang, and W.~Shi,
  ``Checkerboard artifact free sub-pixel convolution: A note on sub-pixel
  convolution, resize convolution and convolution resize,'' \emph{arXiv
  preprint arXiv:1707.02937}, 2017.

\bibitem{Shi_2016_CVPR}
W.~Shi, J.~Caballero, F.~Huszar, J.~Totz, A.~P. Aitken, R.~Bishop, D.~Rueckert,
  and Z.~Wang, ``Real-time single image and video super-resolution using an
  efficient sub-pixel convolutional neural network,'' in \emph{Proceedings of
  the IEEE Conference on Computer Vision and Pattern Recognition (CVPR)}, June
  2016.

\bibitem{toutounchi2019advanced}
F.~Toutounchi and E.~Izquierdo, ``Advanced super-resolution using lossless
  pooling convolutional networks,'' in \emph{2019 IEEE Winter Conference on
  Applications of Computer Vision (WACV)}.\hskip 1em plus 0.5em minus
  0.4em\relax IEEE, 2019, pp. 1562--1568.

\bibitem{nah2017deep}
S.~Nah, T.~Hyun~Kim, and K.~Mu~Lee, ``Deep multi-scale convolutional neural
  network for dynamic scene deblurring,'' in \emph{Proceedings of the IEEE
  conference on computer vision and pattern recognition}, 2017, pp. 3883--3891.

\bibitem{lim2017enhanced}
B.~Lim, S.~Son, H.~Kim, S.~Nah, and K.~Mu~Lee, ``Enhanced deep residual
  networks for single image super-resolution,'' in \emph{Proceedings of the
  IEEE conference on computer vision and pattern recognition workshops}, 2017,
  pp. 136--144.

\bibitem{jain2017patchSR}
S.~Jain, D.~M. Sima, F.~Sanaei~Nezhad, G.~Hangel, W.~Bogner, S.~Williams,
  S.~Van~Huffel, F.~Maes, and D.~Smeets, ``Patch-based super-resolution of mr
  spectroscopic images: application to multiple sclerosis,'' \emph{Frontiers in
  neuroscience}, vol.~11, p.~13, 2017.

\bibitem{he2015delving}
K.~He, X.~Zhang, S.~Ren, and J.~Sun, ``Delving deep into rectifiers: Surpassing
  human-level performance on imagenet classification,'' in \emph{Proceedings of
  the IEEE international conference on computer vision}, 2015, pp. 1026--1034.

\bibitem{perez2020torchio}
F.~P{\'e}rez-Garc{\'\i}a, R.~Sparks, and S.~Ourselin, ``Torchio: a python
  library for efficient loading, preprocessing, augmentation and patch-based
  sampling of medical images in deep learning,'' \emph{arXiv preprint
  arXiv:2003.04696}, 2020.

\bibitem{wang2004imageSSIM}
Z.~Wang, A.~C. Bovik, H.~R. Sheikh, and E.~P. Simoncelli, ``Image quality
  assessment: from error visibility to structural similarity,'' \emph{IEEE
  transactions on image processing}, vol.~13, no.~4, pp. 600--612, 2004.

\bibitem{wang2002universal}
Z.~Wang and A.~C. Bovik, ``A universal image quality index,'' \emph{IEEE signal
  processing letters}, vol.~9, no.~3, pp. 81--84, 2002.

\bibitem{solowij2017chronic}
N.~Solowij, A.~Zalesky, V.~Lorenzetti, and M.~Y{\"u}cel, ``Chronic cannabis use
  and axonal fiber connectivity,'' in \emph{Handbook of Cannabis and Related
  Pathologies}.\hskip 1em plus 0.5em minus 0.4em\relax Elsevier, 2017, pp.
  391--400.

\bibitem{garyfallidis2014dipy}
E.~Garyfallidis, M.~Brett, B.~Amirbekian, A.~Rokem, S.~Van Der~Walt,
  M.~Descoteaux, and I.~Nimmo-Smith, ``Dipy, a library for the analysis of
  diffusion mri data,'' \emph{Frontiers in neuroinformatics}, vol.~8, p.~8,
  2014.

\bibitem{jenkinson2012fsl}
M.~Jenkinson, C.~F. Beckmann, T.~E. Behrens, M.~W. Woolrich, and S.~M. Smith,
  ``Fsl,'' \emph{Neuroimage}, vol.~62, no.~2, pp. 782--790, 2012.

\end{thebibliography}

\end{document}